\documentclass[twocolumn,aps,prb]{revtex4}
\usepackage{epsfig}
\usepackage{amssymb,amsbsy}
\begin{document}
\begin{center}
{\bf  Competing Interactions among Supramolecular Structures on
Surfaces}\\
M. Sayar$^{\dag}$, F. J. Solis$^{\dag}$, M. Olvera de la
Cruz$^{\dag}$, S. I. Stupp$^{\dag\ddag*}$\\
$^{\dag}$ Department of Materials Science and Engineering\\
$^{\ddag}$ Department of Chemistry,$^{\ddag}$ Medical School\\
Northwestern University, Evanston, Illinois 60208\\
$^{*}$ to whom correspondence should be addressed.\\
\end{center}

The design of non-centrosymmetric structures using copolymers has
been a subject of great recent interest.\cite{stupp1,leibler99}
Non-centrosymmetric bulk copolymer structures have been analyzed using
mean field continuous models.\cite{prost94,leibler} In this communication we determine the
conditions for obtaining non-centrosymmetric structures, 
and how to tune the net macroscopic polarizibility solving a
lattice model exactly. The model describes the self-organization
of nanoaggregates of rodcoil molecules into polar films.
\cite{stupp1,pralle3,pralle2} An important property of these
nanoaggregates is their organization into macroscopic polar
materials when cast from solution and annealed without the
involvement of an external electric field. Transmission electron
micrographs of film cross sections and x-ray scattering
experiments reveal the formation of layered domains with a
head-to-tail polar arrangement.\cite{pralle2} Interestingly, the
measured macroscopic polarization of these films is much less
than one would expect for a monodomain of oriented nanoaggregates.
\cite{pralle3} One possible explanation for the relatively small
macroscopic polarity is cancellation among domains in the bulk of
the film. In this communication we explore the possibility of a
non-randomly oriented microstructure by finding the ground state
of a simple model. The model is constructed to account
for the competing interactions among nanoaggregates. The
non-centrosymmetric structure of the nanoaggregates suggests they
have a net dipole moment, and this leads to dipolar interactions
among them. On the other hand, the enthalpic and entropic factors
associated with contacts between coil and rod portions of
neighboring aggregates suggests Ising-like nearest neighbor
interactions.

In our model the nanoaggregates are represented by dipoles of
constant strength $D$ on a cubic lattice of size $a$. While it is
possible to consider very general models in which dipoles are
free to orient in any direction, based on experiments the system
we are interested in modeling contains dipoles than can be
aligned either parallel or anti-parallel to the $z$ axis,
perpendicular to the $x-y$ substrate plane. Also, even though we
included a penalty for reversing the orientation of a dipole with
respect to its neighbors by means of an Ising coupling of
strength $J$, there should be other steric forces associated with
the shape and size of the nanoaggregates which are not considered
here explicitly. These forces could maintain the dipoles oriented
along a given axis. The ground states of the dipole ($J=0$) and
of the Ising ($D=0$) interactions along the $z-x$ plane are shown
in Figures 1.a and 1.b, respectively. The ground state when $J=0$
is antiferromagnetic along $x$ and $y$ (columns along $z$), and
when $D=0$ is a homogeneous state with all dipoles pointing in the
same direction (monodomain). With nonzero values of $D$ and $J$,
an intermediate stripe state of periodicity $\lambda$ is possible
in which dipoles have the same orientation within equal size
domains ($\lambda_{\uparrow}=\lambda_{\downarrow}=\lambda /2$)
along the $x$ or $y$ axis and contiguous domains have
anti-parallel orientations (see Figure 1c). This stripe structure
has been obtained previously in 2d magnetic systems
\cite{whitehead95} and in lipid monolayers
\cite{joanny87,Singer00}. We extended the model to finite
thickness films.\cite{longpaper} In infinite thickness films, we
found a first order transition from a monodomain
($\lambda/a=\infty$) to anti-parallel domains with extremely
small domain sizes ($\lambda/a \leq 4$) as $D/J$ increases.
However, in films of finite thickness, the domain periodicity
$\lambda$ was found to decrease continuously as $D/J$ increases.
We add here a short range interaction between the dipoles and the
surface and show that in this case the dipole up
($\lambda_{\uparrow}$) and dipole down ($\lambda_{\downarrow}$)
domains no longer have equal widths as shown in Figure 1c,
leading to net macroscopic polarization. We find that $\lambda$ and the
macroscopic polarization can be modulated by varying the
film thickness.

\begin{figure} [h]
\epsfig{figure=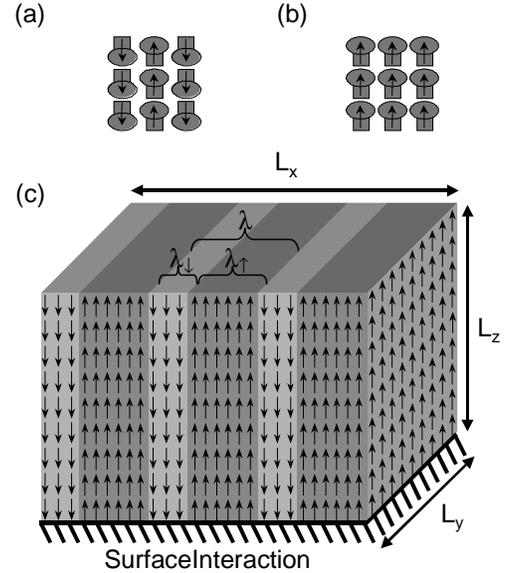,scale=0.8}
\caption{Ground state configurations for a system of supramolecular
aggregates with dipole-dipole interactions only a), and with
Ising interactions only b). The stripe structure under the surface 
influence c).}
\label{fig-01}
\end{figure}

Consider a 3d lattice, with dimensions $L_x$, $L_y$ and $L_z$,
composed of dipoles ${\bf s}_{\bf m}$ with orientations up ${\bf
s}_{\bf m}=(0,0,1)$ or down ${\bf s}_{\bf m}=(0,0,-1)$. The
energy from the Ising interactions is given by,
\begin{equation}
\label{eq:nn} E_{I}=-\frac {J} {2}\sum_{\bf <m, m'>} {\bf
s}_{\bf m}\cdot {\bf s}_{\bf m'}
\end{equation}
where $\bf <m, m'>$ are the nearest neighbors. The
energy due to dipolar interactions is
\begin{equation}
\label{eq:dip}
E_{D}=\frac {D} {2}
    \sum_{\bf m}
        \sum_{{\bf m}\neq{\bf m'}}
        \frac{
            {\bf s}_{\bf m}\cdot {\bf s}_{\bf m'}-3
                ({\bf s}_{\bf m }\cdot \hat{r})
                    ({\bf s}_{\bf m '}\cdot\hat{r})}
            {|{\bf m} - {\bf m'}|^3}
\end{equation}
where $\hat{r}$ is the unit vector in the direction of ${\bf
m}-{\bf m'}$.

The monolayer ($L_z=1$) case can be analyzed readily given that
the second term in $E_{D}$ vanishes. The existence of
anti-parallel domains (stripes) in this 2d case has been
justified by expanding the free energy in powers of the Fourier
components $\phi(\bf{k})$ of a continuous field $\phi(\bf{r})$
proportional to the local polarizability as,\cite{joanny87} $ F =
\sum_{\bf{k}} \phi({\bf{k}}) (G_0^{I} + G_0^{D}) \phi(-\bf{k}) +$
{\it quartic local terms}. Here, the Ising contribution $G_{0}^{I}$ is equal to 
$G_{0}^{I} = 4 \gamma J(-1+ 2a^2k^{2}) + k_BT$, where $\gamma$ is
the number of nearest neighbors, $T$ is temperature and $k_B$ is 
the Boltzman constant. The dipolar contribution is
given by the 2d Fourier transform of the potential in eq
2, $G_{0}^{D}(k) = (2 \pi D /a) (
{_1}F_{2}(-1/2;1/2,1;-a^2k^2/4) - |k|a)$, where ${_1}F_{2}$ is the generalized hypergeometric function.
When $D/J$ is small the $-|k|$ term in $G_0^{D}(ka \ll 1)=(2 \pi
D /a)(1 - |k|a + k^2a^2/2)$  added to the $+k^2$ terms in $G_{0}$
gives a minimum $F$ at a $k^*$ mode that changes continuously
from $k^*=0$ (a monodomain) at $D/J=0$ to $k^*\neq 0$ (a periodic
structure) as soon as $D/J \neq 0$.  The lowest energy structure
is a stripe \cite{alexander,brazovskii} along
$x$ or $y$ with periodicity $\lambda /a= 2\pi/k^*a= 4\pi +
16a/(D/J)$. This continuous analysis, however, is not accurate for neither
$D/J>0.1$ nor low $T$.\cite{Singer00} The stripe structure melts via
defects,\cite{Singer00} with a local $\lambda$ given by the exact $T=0$
value,\cite{whitehead95} $\lambda= A \exp(2aJ/D)$. The
exponential decay of $\lambda/a$ from $\infty$ to $2$ as $D/J$
increases (followed by an antiferromagnetic state), agrees well
with the exact numerical $T=0$ results given below.

In order to analyze the stripe structure in films of arbitrary
thickness we consider the system as a set of $y-z$ planes of one
lattice thickness, stacked along $x$. Since within each plane all
the dipoles have the same orientation, one can assign a new
parameter $s_i^p$ to represent the configuration of all the
dipoles within the plane. The Ising and dipolar energies per
dipole between two planes for a repeat unit of width $\lambda$ is
given by
\begin{equation}
\label{totalnear} E_{I}=-J\frac {\lambda -4} {\lambda}
\end{equation}
and
\begin{equation}
\label{totaldipole} E_{D}=\frac {D} {\lambda}
\sum_{i=1}^{\lambda}\sum_{i'=i+1}^{\infty}V_{D}^p(|i'-i|) s^p_i
s^p_{i'}
\end{equation}
respectively, where $V_{D}^p(|i'-i|)$ is the dipolar potential
between planes $i$ and $i'$. We compute $V_{D}^p(|i'-i|)$ by
summing over all the individual dipole-dipole interactions that
form the planes,
\begin{equation}
\label{eq:planedip}
V_D^p(|i'-i|)=
    \sum_{\bf m_{i}}
        \sum_{\bf m_{i'}}
        \frac{
            {\bf s}_{\bf m_{i}}\cdot {\bf s}_{\bf m_{i'}}-3
                ({\bf s}_{\bf m_{i} }\cdot \hat{r})
                    ({\bf s}_{\bf m_{i'}}\cdot\hat{r})}
            {|{\bf m_{i}} - {\bf m_{i'}}|^3}
\end{equation}
where $\bf m_{i}$ and $\bf m_{i'}$ are the spins in planes $i$ and $i'$ 
respectively. $V_D^p$ has a strong dependence on $L_z$, such that as $L_z$ 
increases $V_D^p$ decreases at short distances and increases at long 
distances. Therefore the range of interaction is longer in thicker films.

The interaction with the substrate can be easily included into
this model. Let us assume the presence of a substrate in the
$x-y$ plane which favors the up configuration for the first layer of dipoles. The energy per dipole due to the interaction with this
substrate can be written as,
\begin{equation}
\label{totalsubstrate} E_{S}=-\frac {S} {\lambda L_{z}}
\sum_{i=1}^{\lambda} (s^p_i)
\end{equation}

\noindent This potential distorts the relative width of domains
of dipoles up and down, such that
$\lambda_{\uparrow}/\lambda \neq 0.5$. 
\begin{figure} [h]
\epsfig{figure=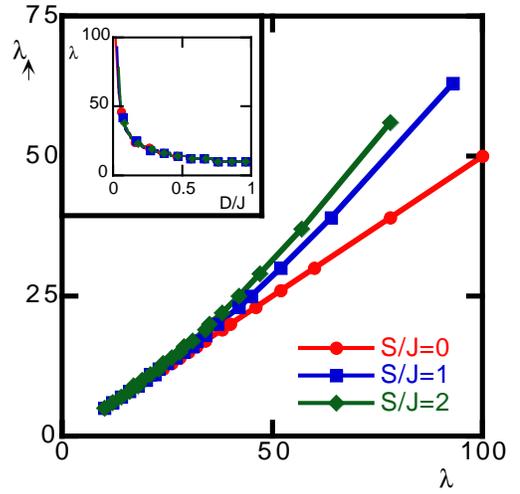,scale=0.8}
\caption{Results for $\lambda_{\uparrow}$ vs. $\lambda$ for 
different surface interaction strengths for a film of 21 layers, 
$S/J=0$ ($\bullet$), S/J=1($\blacksquare$) and $S/J=2$($\blacklozenge$). 
The inset shows $\lambda$ vs. $D/J$ for the same $S/J$ values.}
\label{fig-02}
\end{figure}

Figure 2 shows the effect of surface interaction on the ground
state configuration, obtained by numerically minimizing the total
energy of the system. The calculations are carried out for a film
of 21 layers of dipoles and infinite dimensions in the $\hat{x}$
and $\hat{y}$ directions. Surfaces with $S/J=0,1$ and $2$ give
similar $\lambda$ variations as $D/J$ increases. The domain size
decreases continuously from infinite to finite values over a very
narrow $D/J$ range as shown in the inset in Figure 2. The effect
of the surface interaction is revealed in the
$\lambda_{\uparrow}$ vs. $\lambda$ plot. The $S/J=0$ case yields
$\lambda_{\uparrow}/\lambda=0.5$. As the surface interaction is
turned on, the relative width of up and down domains changes such
that $\lambda_{\uparrow}/\lambda\geq 0.5$ (Figure 1c) and
increase of the surface potential raises this difference further.

The thickness of the film has a drastic effect on the ground state
configuration, such that in the extreme limit of a bulk system
the domain structure is destroyed. We analyzed three systems with
film thickness, $L_z=1$ (a monolayer), $L_{z}=21$ and
$L_{z}=201$, all with $S/J=1$. As shown in Figure 3, these three
systems show a remarkably different parameter dependence. For
large values of $D/J$ the monolayer has a periodicity of
$\lambda/a = 2$, and the domain size grows rapidly at about $D/J
\cong 0.45$, as shown in the inset in Figure 3. This sharp onset
disappears for films with 21 and 201 layers. When we compare the ratio 
$\lambda_{\uparrow}/\lambda$, it is clear that the
monolayer acquires a monodomain configuration with full macroscopic 
polarization and the amount of
polarization decreases as the film thickness increases. Since the
model assumes that the surface interacts only with the first
layer of dipoles, it is clear that the overall effect of the
surface becomes less dominant as $L_{z}$ increases.
\begin{figure} [h]
\epsfig{figure=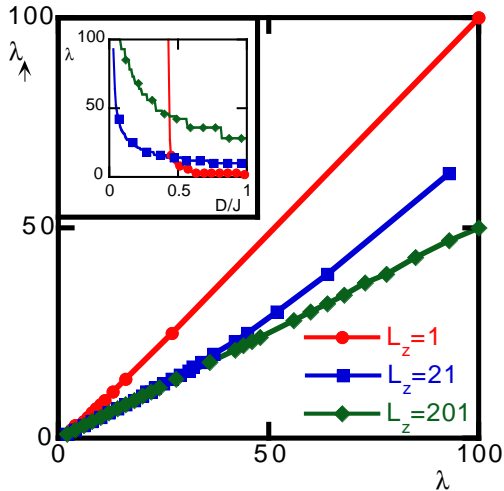,scale=0.8}
\caption{The effect of film thickness $L_{z}$ on the
ratio $\lambda_{\uparrow}/\lambda$ for a monolayer
($\bullet$), 21 layers ($\blacksquare$) and 201 layers ($\blacklozenge$). 
The inset shows $\lambda$ vs. $D/J$ for the same film thicknesses.}
\label{fig-03}
\end{figure}

In our model we assume an Ising interaction which favors parallel
configuration both along the $z$ axis and within the $x-y$ plane.
Based on short range energetic considerations, one might conclude
that the configuration along $z$ would be anti-parallel, since
this would enable to some degree the phase separation of coil and
rod blocks. However this simplification in our model alters the
ground state of the system only when $D/J$ is very small. That
is, if we include an Ising interaction favoring an anti-parallel
configuration along the $z$ axis, the ground state is a
monodomain of bilayers when $D/J \ll 1$. When $D/J$ increases,
however, the parallel configuration along $z$, favored by the
dipolar interaction, dominates even when the Ising interaction
favors an anti-parallel configuration along $z$.

We conclude that the ground state of a system composed of dipolar
supramolecular aggregates that interact with a surface is a
periodic domain structure with a net macroscopic polarization.
Furthermore, one can control the magnitude of this polarization
through variations in the dimension of the system and the
strength of the surface force constant. Monte Carlo simulations for some hexagonal lattices and smectic-A structures with directional order also give stripe structures with similar $\lambda$ dependence on film thickness.

\begin{acknowledgments}
This work was funded by National Science Foundation Grant to S. I. Stupp 
(DMR-9996253) and to M. Olvera de la Cruz (DMR-9807601, and DMR-9632472).
\end{acknowledgments}

\end{document}